\documentclass[preprint,superscriptaddress]{revtex4-2}

\usepackage{times}
\usepackage{graphicx}
\usepackage{amsmath}
\usepackage{amssymb}
\usepackage[colorlinks,linkcolor=blue,citecolor=blue,urlcolor=blue,hyperindex,pdfstartview=FitH,plainpages=false]{hyperref}
\usepackage{float}
\usepackage{lineno}
\DeclareUnicodeCharacter{2212}{\textendash}
\newcommand{\KVS}{KV$ _3$Sb$ _5$}
\newcommand{\CVS}{CsV$ _3$Sb$ _5$}
\newcommand{\RVS}{RbV$ _3$Sb$ _5$}
\newcommand{\AVS}{$A$V$ _3$Sb$ _5$}
\newcommand{\uJcm}{$\mu$J/cm$^2$}
\newcommand{\Tc}{$T_\text{c}$}
\newcommand{\Fc}{$F_\text{c}$}
\newcommand{\Mbar}{$\mathrm{\overline{M}}$}
\newcommand{\bfMbar}{$\mathbf{\overline{M}}$}
\newcommand{\Gammabar}{$\mathrm{\overline{\Gamma}}$}
\newcommand{\Kbar}{$\mathrm{\overline{K}}$}

\begin{document}


\title{Fluctuated lattice-driven charge density wave far above the condensation temperature in kagome superconductor KV$_3$Sb$_5$}
\author{Haoran Liu}
\affiliation{Key Laboratory of Artificial Structures and Quantum Control (Ministry of Education), School of Physics and Astronomy, Shanghai Jiao Tong University, Shanghai 200240, China}
\affiliation{Beijing National Laboratory for Condensed Matter Physics, Institute of Physics, Chinese Academy of Sciences, Beijing 100190, China.}
\author{Shaofeng Duan}
\email{sfduan@iphy.ac.cn}
\affiliation{Beijing National Laboratory for Condensed Matter Physics, Institute of Physics, Chinese Academy of Sciences, Beijing 100190, China.}
\affiliation{Key Laboratory of Artificial Structures and Quantum Control (Ministry of Education), School of Physics and Astronomy, Shanghai Jiao Tong University, Shanghai 200240, China}
\author{Xiangqi Liu}
\affiliation{School of Physical Science and Technology, ShanghaiTech University, Shanghai 201210, China}
\author{Zhihua Liu}
\affiliation{Key Laboratory of Artificial Structures and Quantum Control (Ministry of Education), School of Physics and Astronomy, Shanghai Jiao Tong University, Shanghai 200240, China}
\author{Shichong Wang}
\author{Lingxiao Gu}
\author{Jiongyu Huang}
\author{Wenxuan Yang}
\author{Jianzhe Liu}
\affiliation{Key Laboratory of Artificial Structures and Quantum Control (Ministry of Education), School of Physics and Astronomy, Shanghai Jiao Tong University, Shanghai 200240, China}
\affiliation{Beijing National Laboratory for Condensed Matter Physics, Institute of Physics, Chinese Academy of Sciences, Beijing 100190, China.}
\author{Dong Qian}
\affiliation{Key Laboratory of Artificial Structures and Quantum Control (Ministry of Education), School of Physics and Astronomy, Shanghai Jiao Tong University, Shanghai 200240, China}
\affiliation{Tsung-Dao Lee Institute, Shanghai Jiao Tong University, Shanghai 200240, China}
\affiliation{Collaborative Innovation Center of Advanced Microstructures, Nanjing University, Nanjing 210093, China}
\author{Yanfeng Guo}
\affiliation{School of Physical Science and Technology, ShanghaiTech University, Shanghai 201210, China}
\affiliation{ShanghaiTech Laboratory for Topological Physics, Shanghai 201210, China}
\author{Wentao Zhang}
\email{wentaozhang@iphy.ac.cn}
\affiliation{Beijing National Laboratory for Condensed Matter Physics, Institute of Physics, Chinese Academy of Sciences, Beijing 100190, China.}
\affiliation{Key Laboratory of Artificial Structures and Quantum Control (Ministry of Education), School of Physics and Astronomy, Shanghai Jiao Tong University, Shanghai 200240, China}
\date {\today}
\begin{abstract}
The kagome material AV$_3$Sb$_5$ exhibits multiple exotic orders, including an unconventional charge density wave (CDW). Elucidating the underlying mechanism behind the CDW transition is crucial for unraveling the complex interactions among these phases. However, the driving force of the CDW remains a topic of debate due to the intertwined interactions among the system's various excitations. Here we investigated the CDW transition in \KVS~by isolating the ultrafast electronic phase transition using time- and angle-resolved photoemission spectroscopy. An ultrafast electronic phase transition was observed at a critical photoexcitation fluence, \Fc, without reduction in CDW lattice-distortion-induced band folding. This folded band persisted up to 150 K under equilibrium heating, well above the CDW condensation temperature of \Tc~= 78 K. Notably, the pump-induced band shifts at \Fc~ were comparable to those caused by thermal effects at \Tc. These findings suggest that in \KVS, a fluctuating lattice-driven in-plane CDW emerges above 150 K, with out-of-plane electronic correlations leading to the 2$\times$2$\times$2 CDW near \Tc, offering key insights into the interplay between the electronic and structural dynamics in AV$_3$Sb$_5$.

\textbf{Keywords: } Kagome supercouductor;  Charge density wave; Electonic structure; TRARPES
\end{abstract}

\maketitle

\section*{Introduction}
The kagome lattice, characterized by its network of corner-sharing triangles, exhibits various novel quantum phenomena, such as quantum magnetism \cite{Zhou2017}, non-trivial topology \cite{Tang2011}, superconductivity \cite{Yu2012, Wang2013, Kiesel2013}, and charge density wave (CDW) \cite{ Wang2013, Kiesel2013}. This unique structure provides an excellent platform for studying these fascinating orders. The recently discovered kagome metal \AVS~($A$ = K, Rb, Cs) \cite{Ortiz2019, Yin2022}, featuring a vanadium-formed kagome lattice (Fig. \ref{Fig1}\textbf{a}), particularly garners significant attention for its unique properties, including topological superconductivity with a pair density wave \cite{Chen2021},  unconventional CDW accompanied by nematicity \cite{Nie2022, Xu2022}, and possible time-reversal symmetry breaking \cite{Xing2024}. The CDW state in \AVS~appears below 78 K for $A$ = K \cite{Ortiz2019}, 103 K for $A$ = Rb \cite{Yin2021}, and 94 K for $A$ = Cs \cite{Ortiz2020}. The CDW  partially gaps the Fermi surface \cite{Liang2021, Luo2022}, and an unusual competition with superconductivity was revealed in the recent investigations under moderate pressure \cite{Yu2021, Chen2021a}.

The interplay between CDW, superconductivity, and other exotic phases highlights the importance of understanding the mechanism of CDW in \AVS~to fully grasp its complex phase diagram. Despite numerous experimental and theoretical investigations into the CDW transition, no consensus on its origin has been reached so far due to complex interactions in \AVS.  For instance, some studies claimed that the CDW is driven by Fermi surface nesting, arising from parallel hexagon Fermi surface and van Hove singularities (VHSs) with a high density of states (DOS) near the Brillouin zone boundary, as evidenced by angle-resolved photoemission spectroscopy (ARPES) \cite{Cho2021, Kang2022}, optical measurements \cite{Zhou2021, Yu2023}, and density functional theory calculations \cite{Tan2021, Christensen2021}. Conversely, inelastic scattering experiments do not detect a soft acoustic phonon mode \cite{Xie2022, Li2021}, challenging the traditional nesting picture. An alternative hypothesis, suggesting a first-order CDW transition driven by electron-phonon coupling, is supported by the calculations of electronic and phononic response functions \cite{Wang2022, Si2022, Ye2022}, electron-phonon coupling effects measured by Raman spectroscopy \cite{Liu2022, He2024}, and kinks due to electron-boson interactions observed in ARPES experiments \cite{Luo2022}. Consequently, it is essential to establish a comprehensive understanding of the intertwined interactions in \AVS~and address these conflicts from a new perspective. 

Ultrafast light-matter interaction is a promising route to elucidate the microscopic origins of the phase transition in correlated systems. With optical excitation, ultrafast laser pulses can isolate different degrees of freedom in the time domain \cite{Johnson2017} and lead to purely electronic phase transitions without disturbing lattice structure within hundreds of femtoseconds \cite{Rohwer2011, Yang2022a, Tang2020}. In contrast to the equilibrium transition simultaneously involving electronic and structural degrees of freedom, the light-induced purely electronic phase transition primarily reflects the response of electronic states after photoexcitation. Upon comparing the transient evolution of electronic states induced by optical pumping with those resulting from the equilibrium thermal effects, it becomes possible to determine the individual contributions of electronic and structural degrees of freedom on the phase transition \cite{Yang2022a}. This proposal can be achieved through a combination of equilibrium and nonequilibrium measurements, such as time-resolved ARPES (TRARPES), which is a powerful experimental tool for measuring both transient and equilibrium electronic structures \cite{Boschini2024}, providing a new method to elucidate the mechanism of the CDW transition in \AVS.

Herein, we conducted TRARPES measurements to investigate the origin of the CDW phase transition in a \AVS~candidate with $A$ = K, in which two equilibrium phase transitions and one purely electronic phase transition related to the CDW transition were identified. Detailed temperature-dependent electronic structure measurements demonstrated two equilibrium transitions with one at T$^*$ above 150 K and another at \Tc~= 78 K, corresponding to the preformation of in-plane CDW and condensation of three-dimensional CDW, respectively. Fluence-dependent energy shift of $\gamma$ band and the reduction of lattice-distortion-induced energy gap at -0.6 eV at \Mbar~point revealed a critical fluence at \Fc~$\approx$ 75 \uJcm, associated with the breakdown of out-of-plane CDW order.
However, there were no significant changes in the CDW-induced folded band and coherent phonon modes at the highest pump fluence we studied, indicating that the optical excitation alone cannot destroy the CDW-coupled lattice distortions. Additionally,  the comparable energy shifts induced by optical excitation at \Fc~and equilibrium thermal effect at \Tc~suggest that the electronic correlations play a crucial role in the formation of three-dimensional CDW at \Tc. 
The remaining order with folded bands and energy gap opening persists up to T$^*$ possibly resulting from the fluctuated lattice-driven in-plane CDW order. Our work provides important insights into the role of electronic correlations and electron-phonon coupling in the formation of symmetry-breaking states.

\section*{Results}
\begin{figure} [htbp]
\centering\includegraphics[width = 1\columnwidth] {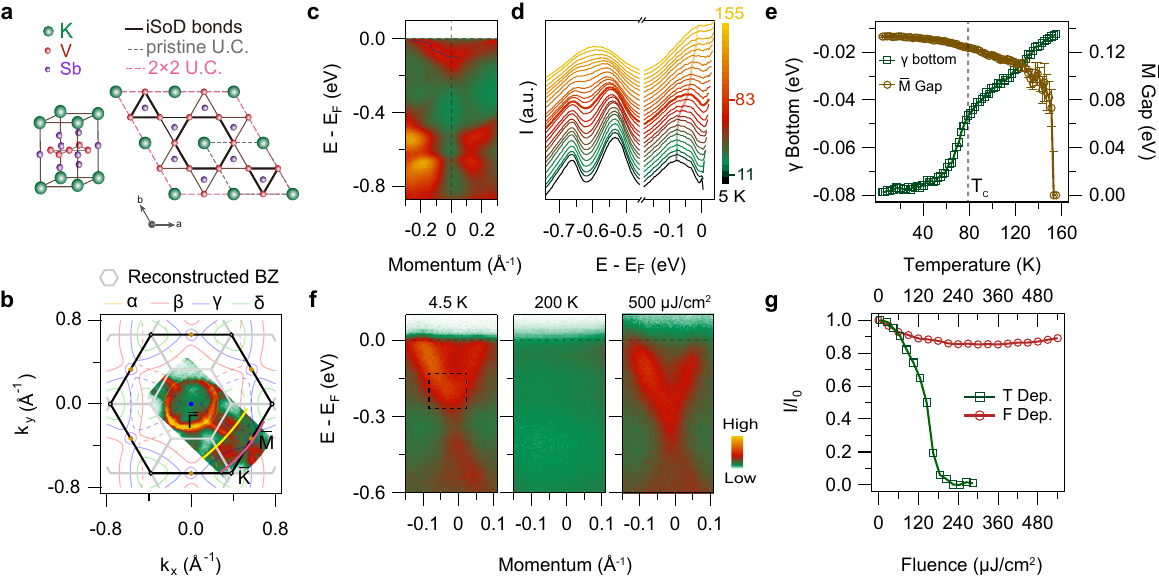}
\caption{\textbf{Temperature- and fluence-dependent photoemission spectra in \KVS.} \textbf{a}, Crystal structure of \KVS. After the CDW transition, the kagome lattice undergoes iSoD distortion. \textbf{b}, Theoretical and measured Fermi surface contour.  The measured Fermi surface was integrated within $\pm$10 meV near the Fermi energy at 4.5 K. The theoretical Fermi surface (adapted from Ref. \cite{Luo2022}) is shown with colored curves representing various bands ($\alpha$, $\beta$, $\gamma$, $\delta$) and dashed curves indicating folded bands. \textbf{c}, ARPES spectra taken along \Kbar -\Mbar -\Kbar~ (pink momentum cut in \textbf{b}) at 4.5 K measured using a 7 eV laser. The gray curve represents $\gamma$ band dispersion near \Mbar, obtained from the peak of EDCs. \textbf{d}, EDCs at \Mbar~($k$ = 0 \AA$^{-1}$ in \textbf{c}) divided by Fermi-distribution function at equilibrium temperatures from 5 to 155 K. \textbf{e}, Temperature-dependent energy shift of the $\gamma$ band bottom and size of the \Mbar G. \textbf{f}, ARPES spectra of the folded band (yellow momentum cut in \textbf{b}) at 4.5 K, 200 K, and 4.5 K with a pump fluence of 500 \uJcm~at a delay time of 0.2 ps. \textbf{g}, Integrated intensities of the folded band (square area in \textbf{f}) as a function of temperature and fluence.}
\label{Fig1}
\end{figure}

\KVS~has a layered crystal structure with space group P6/mmm (No. 191), characterized by a typical kagome structure of a vanadium net. The in-plane hexagonal lattice constant $a$ is 5.482 \AA, and the out-of-plane lattice constant $c$ is 8.948 \AA~\cite{Ptok2022}. In the CDW state, the vanadium kagome net exhibits a 2$\times$2 inverse star-of-David (iSoD) distortion (Fig. \ref{Fig1}\textbf{a}) \cite{Kang2023}. The Fermi surface topology of \KVS~(Fig. \ref{Fig1}\textbf{b}), consists of a circular pocket ($\alpha$) and a large hexagon-shaped sheet ($\beta$) centered around \Gammabar, and triangular pockets ($\gamma$ and $\delta$) around \Kbar~\cite{Luo2022}. The $\alpha$ pocket arises from Sb $5p$ orbitals, whereas the $\beta$, $\gamma$, and $\delta$ pockets originate from V $3d$ orbitals. The theoretical Fermi surface is consistent with our photoemission measurement at 4.5 K in the CDW state,  revealing clear folded bands primarily from one of the CDW wavevectors, which possibly resulted from the nematicity \cite{Jiang2023} or the matrix element effect in the photoemission experiment \cite{Luo2022}. The shallow $\gamma$ band along the  \Kbar-\Mbar-\Kbar~direction at 4.5 K exhibits strong photoemission intensity, attributed to the VHSs near the M/L point, and shows an obvious kink feature at around 30 meV in the electronic dispersion (Fig. \ref{Fig1}\textbf{c}, Supplementary Fig. 2).
An energy gap (\Mbar G)  in the $\delta$ band is situated around -0.6 eV away from the Fermi level at \Mbar~point, and it is a result of CDW-induced lattice distortion, consistent with prior study \cite{Luo2022}.

To study the equilibrium CDW transition in electronic structure, we performed detailed temperature-dependent measurements using an advanced sample-position autocorrection system. This system enabled us to consistently track the equilibrium electronic structure at the same spot with a precision of better than 1 $\mu$m over a wide temperature range \cite{Duan2022}. As the temperature increases, the $\gamma$ band bottom shifts dramatically upwards when approaching \Tc~ (Fig. \ref{Fig1}\textbf{d}) and undergoes a striking transition with total energy shifts of about 32 meV at \Tc~compared with its energy at the lowest temperature (Fig. \ref{Fig1}\textbf{e}).
Meanwhile, the dispersion of the $\gamma$ band becomes flatter as the temperature approaches \Tc, suggesting an effective mass modification through the CDW transition (Supplementary Fig. 3\textbf{a} and Note 2). In comparison, the energy difference of the two peaks in energy distribution curves (EDCs) at \Mbar G (Fig. \ref{Fig1}\textbf{d}) shows a slight decrease of only about 13 meV without a clear transition feature at \Tc~(Fig. \ref{Fig1}\textbf{e}). However, the intensity within the gap shows a noticeable transition at \Tc~(Supplementary Fig. 4), suggesting a change in the rate of gap filling associated with the CDW transition. The electronic structure at \Mbar G subsequently merges into a single peak at approximately 150 K, well above \Tc~(Fig. \ref{Fig1}\textbf{d}), indicating that CDW-coupled structural distortion persists at least up to 150 K. These observations are consistent with the fact that the lattice-distortion induced folded band can be evidenced at 155 K (Supplementary Fig. 5).
The intensity of the folded band vanishes above around 180 K (Figs. \ref{Fig1}\textbf{f} and \ref{Fig1}\textbf{g}), possibly due to the potassium desorption from the sample surface at high temperatures, since the folded bands do not reemerge upon cooling back to low temperatures, and this aligns with the electronic structure observed on the Sb-rich termination \cite{Jiang2023, Kato2022}. Remarkably, the folded band remains intact under optical excitation with a delay time of 0.2 ps at the maximum fluence of 500 \uJcm~in our study (Fig. \ref{Fig1}\textbf{g}). At this fluence, the estimated nonequilibrium electronic temperature is around 500 K. This observation is in contrast to the equilibrium electronic structures with disappeared band folding above the temperature of 180 K, suggesting that the folded band at high temperature is driven by the structural factor and the lattice distortion remains isolated from the low-temperature electronic states within hundreds of femtoseconds. 

\begin{figure}[htbp]
\includegraphics[width = 0.5\columnwidth] {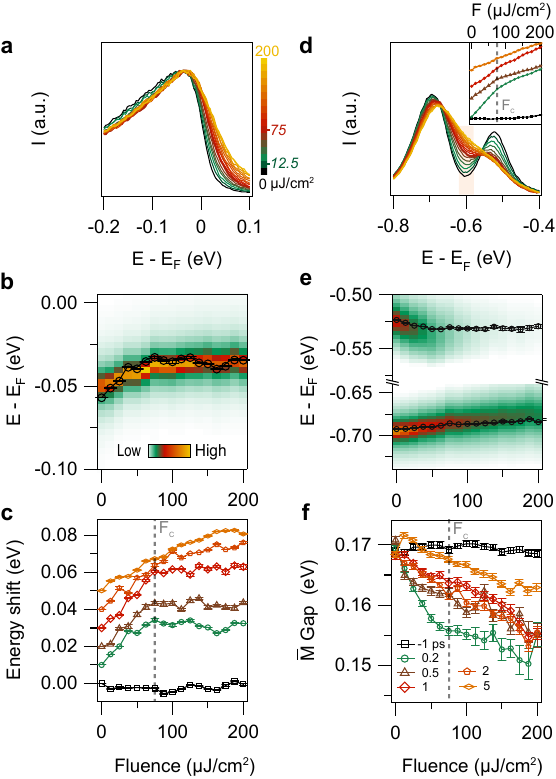}
\centering\caption{\textbf{Fluence-dependent spectra at \bfMbar~at a base temperature of 4.5 K.} \textbf{a}, Fluence-dependent EDCs at 0.2 ps after photoexcitation. EDCs are normalized to the same height. \textbf{b}, Second-derivative image of the fluence-dependent intensity spectrum from \textbf{a}. \textbf{c}, Fluence-dependent energy shifts of the $\gamma$ band bottom at delay times of -1, 0.2, 0.5, 1, 2, and 5 ps, with applied energy offsets. The band shifts were determined from the left edge shift of the peak to minimize the influence of the pump-induced space charge effect and the broadening due to the increased electronic temperature. \textbf{d}, Fluence-dependent EDCs near the \Mbar G at 0.2 ps. Inset: Fluence-dependent integrated intensity within the \Mbar G (shaded area in \textbf{d}) at delay times of -1, 0.2, 0.5, 1, and 5 ps. \textbf{e}, Second-derivative image of the fluence-dependent intensity spectrum from \textbf{d}. \textbf{f}, Fluence-dependent size of the \Mbar G at delay times of -1, 0.2, 0.5, 1, 2, and 5 ps.}
\label{Fig2}
\end{figure}

The CDW electronic states can be effectively driven by optical excitation, while the pump pulse cannot diminish the lattice distortion. 
Fluence-dependent measurements show that the $\gamma$ band bottom shifts towards the Fermi level at low fluences, saturating by about 30 meV at a critical fluence of \Fc~= 75 \uJcm~at delay times of 0.2, 0.5, 1 ps (Fig. \ref{Fig2}\textbf{c}), and exhibiting a broader Fermi edge at higher fluences (Figs. \ref{Fig2}\textbf{a} and \ref{Fig2}\textbf{b}). As optical excitation could transiently enhance the electronic temperature and screen the  Coulomb interaction, we attribute the critical pump fluence (\Fc) to the signature of the destroying the electronic correlations, consistent with the photoinduced nonthermal CDW transition in the ultrafast optical spectroscopy experiments \cite{Ratcliff2021, Wang2021}. 
The absence of a resolvable critical fluence at delay times longer than 2 ps in the band shift suggests that the recovery time of the phase transition is shorter than 2 ps. The two peaks of \Mbar G in the EDCs near -0.6 eV move closer upon enhancing the pump fluence and remain nearly unchanged above the critical fluence \Fc, as shown in Figs. \ref{Fig2}\textbf{d} and \ref{Fig2}\textbf{e}. The \Mbar G is reduced by approximately 13 meV at the critical fluence and remains nearly constant at higher fluences at 0.2 ps after photoexcitation. 
Although the \Mbar G does not exhibit a resolvable critical feature at a longer delay time, the spectral intensity within the gap displays a similar critical pump fluence feature at 0.5 ps and 1 ps (inset of Fig. \ref{Fig2}\textbf{d}), consistent with the observations of the electronic states near Fermi energy.
Interestingly, the energy changes of both the $\gamma$ band bottom and the \Mbar G induced by the photoexcitation at the \Fc~are comparable to the changes observed by thermally heating the sample temperature to \Tc~(Fig. \ref{Fig1}\textbf{e}), which suggests the electronic correlation dominates the CDW transition at \Tc~and will be discussed later on.

\begin{figure}[htbp]
\centering\includegraphics[width = 0.5\columnwidth] {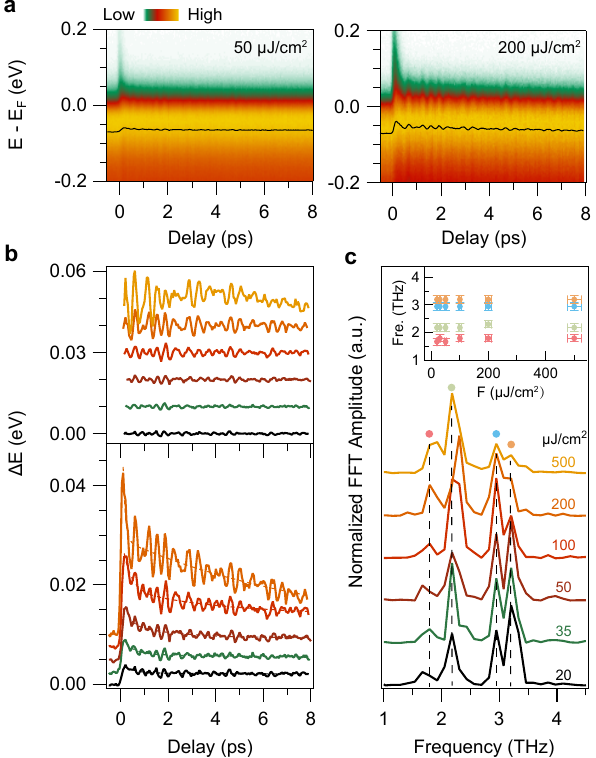}
\caption{\textbf{Pump-induced coherent oscillations of the energy shifts near \bfMbar.} \textbf{a}, Photoemission intensity as a function of energy and delay time at the momentum of the $\gamma$ band bottom with pump fluences of 50 and 200 \uJcm. \textbf{b}, Energy shift of the $\gamma$ band bottom upon pump (bottom) and coherent oscillations after removing the incoherent component (top) at pump fluences of 20, 35, 50, 100, 200, and 500 \uJcm~, with applied energy offsets. Note that the energy shift at 500 \uJcm~is not shown due to the pump-induced space charge effect causing a significant downward shift in the background (Supplementary Fig. 7). \textbf{c}, Fast Fourier transform (FFT) amplitude of the oscillations in \textbf{b}. The FFTs are performed between 0.2 ps and 8 ps, and the FFT amplitudes are normalized to the same height. Inset: Fluence-dependent phonon frequencies of 1.7, 2.3, 2.95, and 3.2 THz, determined from the  FFT results.}
\label{Fig3}
\end{figure}

Typically, the softening of the CDW-coupled amplitude coherent phonon mode at the critical fluence is a signature of the recovery of CDW-induced lattice distortion \cite{MoehrVorobeva2011}.  Examining the CDW-coupled amplitude mode is vital for assessing the influence of the lattice on the CDW transition.
Experimentally, coherent oscillations in the electronic structure near the Fermi level at \Mbar~can be clearly observed both in the raw data and the extracted pump-induced energy shifts as a function of delay time for all photoexcitation fluence up to 500 \uJcm,  significantly exceeding \Fc~(Figs. \ref{Fig3}\textbf{a} and \ref{Fig3}\textbf{b}). Four coherent phonon modes with frequencies of about 1.7, 2.3, 2.95, and 3.2 THz are resolved in the Fourier transformation amplitudes (Fig. \ref{Fig3}\textbf{c}). The 1.7 THz mode closely matches the $E_{2g}$ phonon, which involves the motion of potassium atoms and appears below \Tc~\cite{Wu2022}, indicating it is strongly coupled to the CDW formation. The 3.2 THz mode possibly corresponds to the 3.1 THz mode found in \CVS~via ultrafast optical spectroscopy measurements, which emerges below $\sim$ 60 K and is attributed to another phase transition related to a unidirectional order \cite{Wang2021, Ratcliff2021}. According to the calculation \cite{Ptok2022}, the 2.3 and 2.95 THz modes exhibit large phonon DOS from Sb, regardless of CDW distortion, and have not been reported in previous experiments. Notably, all the phonon modes persist across the entire range of pump fluences, exhibiting only an increase in amplitude and without any signature of frequency softening (Fig. \ref{Fig3}\textbf{c}). This implies that the CDW-coupled lattice distortion cannot be significantly altered by intense light pumping in \KVS. The observed soft modes in \CVS~at much higher fluences by Azoury et al. \cite{Azoury2023}, might result from disrupting the in-plane CDW order, beyond the scope of our study.

\section*{Discussion}

The CDW state manifests as periodic modulations of the charge density and is accompanied by lattice distortions, modifying the electronic structure with gap opening and band folding. The observed folded bands of \KVS~in our experiments originate from the bulk CDW-coupled lattice distortions rather than the cleaved surface reconstruction due to the following reasons. 
First, the electronic structures were measured on the K-rich termination, which is expected to preserve the bulk CDW properties, while the Sb-rich termination does not exhibit CDW properties \cite{Jiang2023, Kato2022, Cai2024}. Previous studies have identified a temperature-independent surface quantum well state in the Sb-rich termination of \CVS~\cite{Cai2024}, which cannot account for the observed results with folded bands and gap opening at high temperatures in our experiment on the K-rich termination. Since the K-terminated surface is a tiny domain on the cleaved surface \cite{Jiang2023}, the probe beam spot is focused on a single K-rich termination surface by utilizing a high-precision sample-position autocorrection system with submicron precision \cite{Duan2022}. This approach enables us to avoid missing the K-terminated surface in the temperature-dependent measurements. Second, the observed folded bands in our study align closely with the bulk CDW wavevector, indicating that the superstructure modulation at the surface resulted from the bulk CDW transition. Although a $\sqrt{3}\times\sqrt{3}$ surface reconstruction of Rb has been found in \RVS~\cite{Yu2022, Meng2023}, this wavevector does not align with the folded bands observed in our experiments (Supplementary Fig. 6). Finally, the observed band folding far above \Tc~cannot be attributed to a surface-enhanced CDW, which possesses the same wavevector as the bulk on the sample surface and has been reported in the CDW material of K$_{0.9}$Mo$_{6}$O$_{17}$ \cite{Mou2016}. Unlike K$_{0.9}$Mo$_{6}$O$_{17}$, we do not identify two distinct sets of CDW bands in \KVS. 

Despite the potential possibility of potassium desorption at high temperatures, the closure of \Mbar G around 150 K and the disappearance of the folded band above 180 K strongly suggest an intrinsic temperature scale (T$^*$) much higher than \Tc. The signature of the temperature scale T$^*$ also appeared in the resistance measurements, as indicated by a resolvable dip between 160 and 200 K (Supplementary Fig. 8).  Additionally, the \Mbar G is an order of magnitude larger than the CDW gap near the Fermi level and does not exhibit a signature of transition at \Tc, indicating it arising from another high-temperature transition. A similar high-temperature scale above 150 K has also been observed in the diffraction experiments in \CVS~and it is regarded as the onset of strong CDW fluctuations \cite{Chen2022a}, corresponding to the emergence of in-plane lattice distortions \cite{Park2023}. Theoretically, the  \Mbar G is the result of the CDW-related iSoD or SoD lattice  \cite{Luo2022, Hu2022a, Kang2023}. The observation of robust folded bands at the highest pump fluence after screening the electronic correlations suggests the structural origin of the lattice distortion. Strong electron-phonon coupling, which induces CDW fluctuation in the material, has been reported \cite{Subires2023, Yang2023}. Consequently, we speculate that structural instability drives the formation of the fluctuated in-plane CDW at high temperatures, as proposed theoretically \cite{Park2023}. 

Furthermore, the ultrafast electronic dynamics in \KVS~suggest that the CDW transition at \Tc~is driven by electronic correlations. With strong optical excitation, the infrared laser pulses induced a purely electronic phase transition via carrier screening without disturbing the lattice structure, as illustrated by the robust CDW-coupled phonon modes in Fig. \ref{Fig3}\textbf{c}.  The photoinduced energy shift of the $\gamma$ band bottom and the decrease in the \Mbar G size at \Fc~were about 30 and 13 meV, respectively, and the critical fluence \Fc~indicates a light-induced breakdown of the electronic correlations. The energy shift of the $\gamma$ band bottom and the reductions of the \Mbar G at \Fc~(Figs. \ref{Fig2}\textbf{c} and \ref{Fig2}\textbf{f}) are comparable to those induced by thermal effects when raising the temperature to \Tc~(Fig. \ref{Fig1}\textbf{e}), suggesting that the electronic correlations dominate the equilibrium CDW transition at \Tc. As the intralayer distortions have existed at higher temperatures, the electronic correlations stabilize the out-of-plane order via interlayer coupling at \Tc. It is consistent with previous reports that the out-of-plane order is more sensitive than the in-plane order by light pumping \cite{Ning2024} and chemical doping \cite{Xiao2023, Kautzsch2023} in \CVS~and that out-of-plane order is crucial for stabilizing the CDW distortion \cite{Park2023, Li2023, Deng2023}. Therefore, the critical fluence \Fc~observed in the light-induced electronic phase transition is a signature of melting the interlayer order. Additionally, VHSs close to the Fermi level have been reported near the M/L point in the momentum space \cite{Cho2021, Hu2022}. The $\gamma$ band bottom, which corresponds to one of these VHSs, shows a substantial shift near \Tc~ (Fig. \ref{Fig1}\textbf{d}), indicating a transition of the electronic effective mass as a function of temperature (Supplementary Fig. 3\textbf{b}). Below \Tc, the $\gamma$ band bottom is pushed away from the Fermi level (Fig. \ref{Fig1}\textbf{c}), consistent with a suppression of electronic DOS near the Fermi level. The aforementioned results suggest that the interlayer electronic correlations may be established through the Fermi surface nesting along the $k_z$ direction to maintain the out-of-plane CDW order.

\begin{figure}[htbp]
\centering\includegraphics[width = 1\columnwidth] {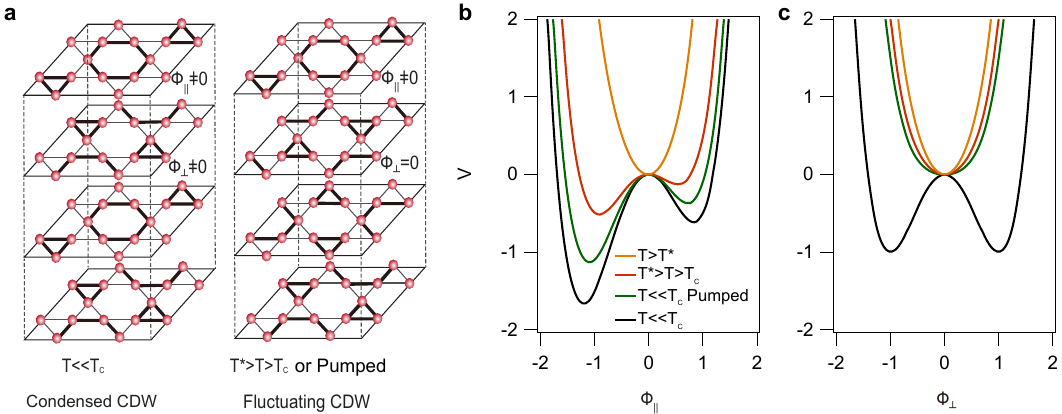}
\caption{\textbf{A Ginzburg-Landau perspective on the CDW phase transition in \KVS~under both equilibrium and nonequilibrium cases.} \textbf{a}, The associated lattice distortions in the condensed and fluctuating CDW phases. The fluctuating state exhibits an iSoD in-plane lattice distortion with an in-plane order parameter of $\Phi_{\parallel}\ne0$ and an out-of-plane order parameter of $\Phi_{\perp}=0$. Below \Tc, the system condenses into a fully developed three-dimensional CDW state with $\Phi_{\parallel}$ and $\Phi_{\perp}\ne0$. \textbf{b}, The Ginzburg-Landau potential as a function of $\Phi_{\parallel}$. At temperatures above T$^*$, $\Phi_{\parallel}=0$. As the temperature drops below T$^*$, $\Phi_{\parallel}$ emerges with strong fluctuations. The asymmetry of the potential results from the energy difference between iSoD and SoD lattice distortions \cite{Tan2021}. The two-well potential becomes steeper as the temperature drops, and the photoexcitation only weakly modifies the in-plane order below \Tc. \textbf{c}, The Ginzburg-Landau potential as a function of $\Phi_{\perp}$. The out-of-plane order parameter of $\Phi_{\perp}$ emerges as the temperature below \Tc, indicating the transition of the system into the condensed CDW state. Upon photoexcitation, the interlayer coherence is destroyed with $\Phi_{\perp}$ = 0.}
\label{Fig4}
\end{figure}

To interpret the experimental observations, we propose a phenomenological model to describe the CDW formation process in \AVS. The symmetry-broken CDW state is determined by the minimum of the energy potential, which is a function of order parameters with in-plane and out-of-plane components, as illustrated in Figure \ref{Fig4}. At temperatures above T$^*$, the free energy landscape is a single well with the kagome plane maintaining its pristine structure. As the temperature decreases below T$^*$, the lattice-driven fluctuated in-plane CDW order begins to develop within each layer (Fig. \ref{Fig4}\textbf{b}) but without interlayer coherence (Fig. \ref{Fig4}\textbf{c}). Since in-plane lattice distortion dominates most of the time (right panel of Fig. \ref{Fig4}\textbf{a}) \cite{Park2023}, ARPES is capable of detecting the folded bands induced by the fluctuating distortion, even though it is a time-averaged experimental technique. As the temperature drops below \Tc, the out-of-plane order emerges (Fig. \ref{Fig4}\textbf{c}) due to electronic correlations, leading to three-dimensional CDW condensation and the formation of the 2$\times$2$\times$2 CDW superstructure (left panel of Fig. \ref{Fig4}\textbf{a}). With strong ultrafast photoexcitation, the transiently excited nonequilibrium quasiparticles could screen the electronic correlation and lead to the destruction of the out-of-plane coherence within hundreds of femtoseconds, while the in-plane order is preserved. Consequently, the comparable energy shifts at \Fc~and \Tc~resulted from the quenching of the out-of-plane CDW.

In summary, we elucidated the dynamics of CDW formation in \KVS~by combining high-resolution time-resolved ARPES and precise temperature-dependent measurements. We found that ultrafast photoexcitation induces an electronic phase transition while preserving the in-plane CDW order. Temperature-dependent measurements reveal that the in-plane CDW order develops and fluctuates at much higher temperatures than \Tc, driven by structural instability. The comparison of the $\gamma$ band shifts and the reductions in \Mbar G size caused by ultrafast photoexcitation and equilibrium thermal effects suggests that electronic correlations primarily drive the condensation of the CDW state at \Tc. These findings provide crucial experimental insights into CDW formation in \AVS, advancing our understanding of the intricate interplay between electronic and structural dynamics in these exotic systems.

\section*{Methods}
The TRARPES measurements were conducted using infrared pump laser pulses with a photon energy of 1.77 eV and a repetition rate of 500 kHz to excite the sample into nonequilibrium states \cite{Yang2019}. Subsequently, ultraviolet probe pulses (6.05 eV) were used to photoemit electrons. The time resolution was achieved by varying the delay between the pump and probe pulses and was optimized to 113 fs for the best performance. The energy resolution was set to 50 meV to balance the space charge effect and improve emission counts. The spot sizes of the pump and probe beams were approximately 80 and 22 $\mu$m, respectively. To minimize the pump-induced space charge effect \cite{Azoury2023}, the pump light polarization was set parallel to the sample surface (Supplementary Note 1). High-resolution ARPES using 7 eV photons was conducted with an energy resolution of 4 meV \cite{Huang2022}. During the temperature-dependent measurements, the sample position was autocorrected with a precision better than 1 $\mu$m \cite{Duan2022}. High-quality \KVS~single crystals were grown using the self-flux method \cite{Yang2023}, and their resistance and Laue pattern were measured (Supplementary Fig. 8). The samples were cleaved under ultrahigh-vacuum conditions with a pressure of $\mathrm{3\times10^{-11}}$ Torr in ARPES measurements. All TRARPES measurements were carried out at an equilibrium temperature of approximately 4.5 K.

\begin{acknowledgements}

W. T. Z. acknowledges support from the National Key R\&D Program of China (Grants No. 2021YFA1401800 and No. 2021YFA1400202) and National Natural Science Foundation of China (Grant No. 12141404) and Natural Science Foundation of Shanghai (Grants No. 22ZR1479700 and No. 23XD1422200). S. F. D. acknowledges support from the China Postdoctoral Science Foundation (Grant No. 2022M722108) and China National Postdoctoral Program for Innovative Talents (Grant No. BX20230216) and National Natural Science Foundation of China (Grant No. 12304178). Y. F. Guo acknowledges the National Key R\&D Program of China (Grant No. 2023YFA1406100) and the Double First-Class Initiative Fund of ShanghaiTech University. D.Q. acknowledges support from the National Key R\&D Program of China (Grants No. 2022YFA1402400 and No. 2021YFA1400100) and National Natural Science Foundation of China (Grant No. 12074248).
\end{acknowledgements} 
\section*{Author Contributions}
W.T.Z. proposed and designed the research. S.F.D., L.X.G., S.C.W., H.R.L., J.Y.H., J.Z.L., and W.T.Z. contributed to the development and maintenance of the TRARPES system. H.R.L., S.F.D., J.Y.H., and W.X.Y. collected the TRARPES data. Z.H.L. performed the transport measurement. W.X. and Y.F.G. prepared the single crystal sample. W.T.Z. wrote the paper with H.R.L., S.F.D., and D.Q. All authors discussed the results and commented on the manuscript.
\section*{Competing Interests}
The authors declare that they have no competing interests.

\section*{Data Availability}

The data that support the plots within this paper and other findings of this study are available from the corresponding author upon reasonable request. Correspondence and requests for materials should be addressed to S.F.D. (sfduan@iphy.ac.cn) and W.T.Z. (wentaozhang@iphy.ac.cn).


%

\newpage

\begin{figure*}[htbp]
\centering\includegraphics[width = 1\columnwidth,page=1] {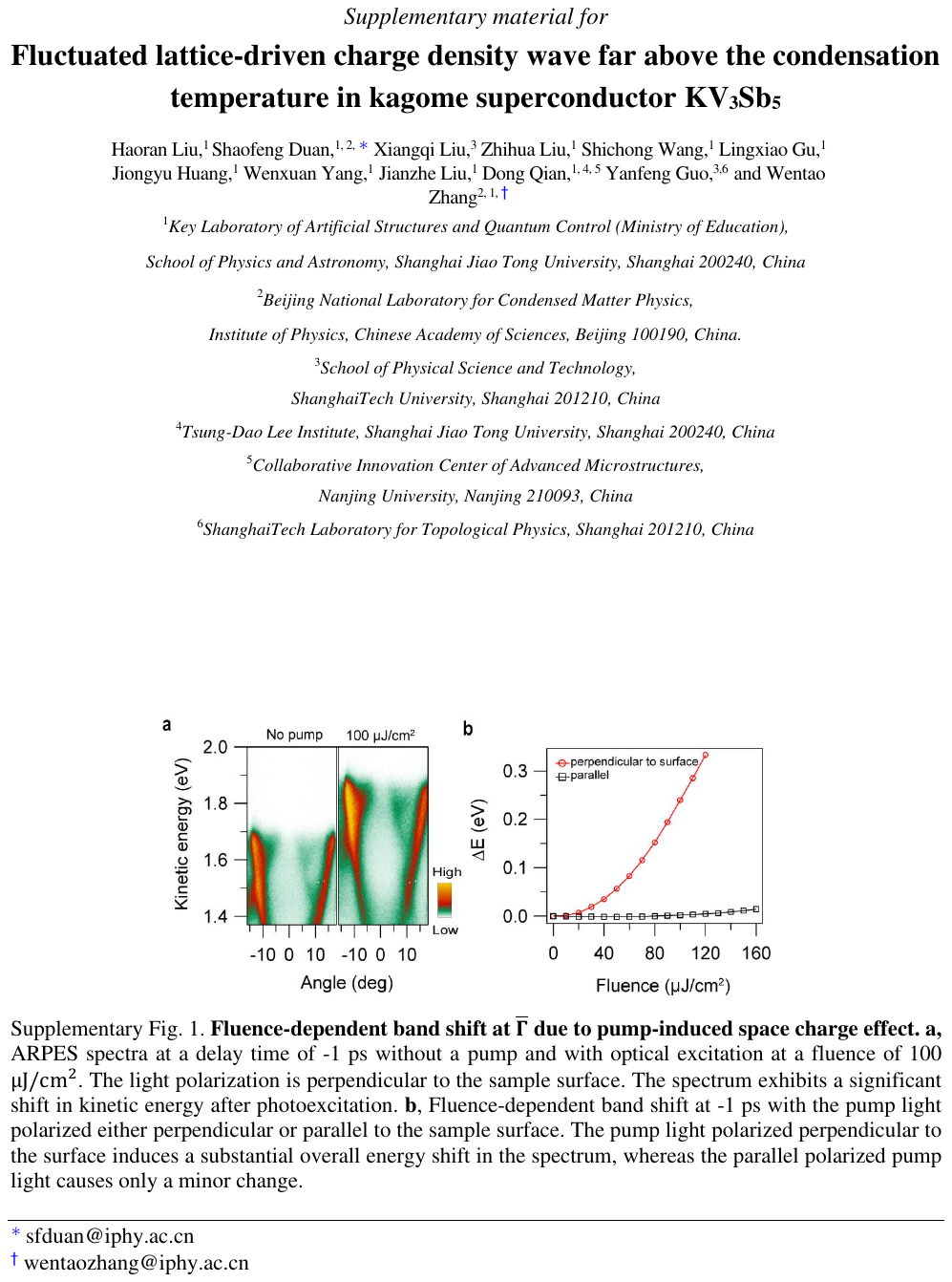}
\end{figure*}
\begin{figure*}[htbp]
\centering\includegraphics[width = 1\columnwidth,page=2] {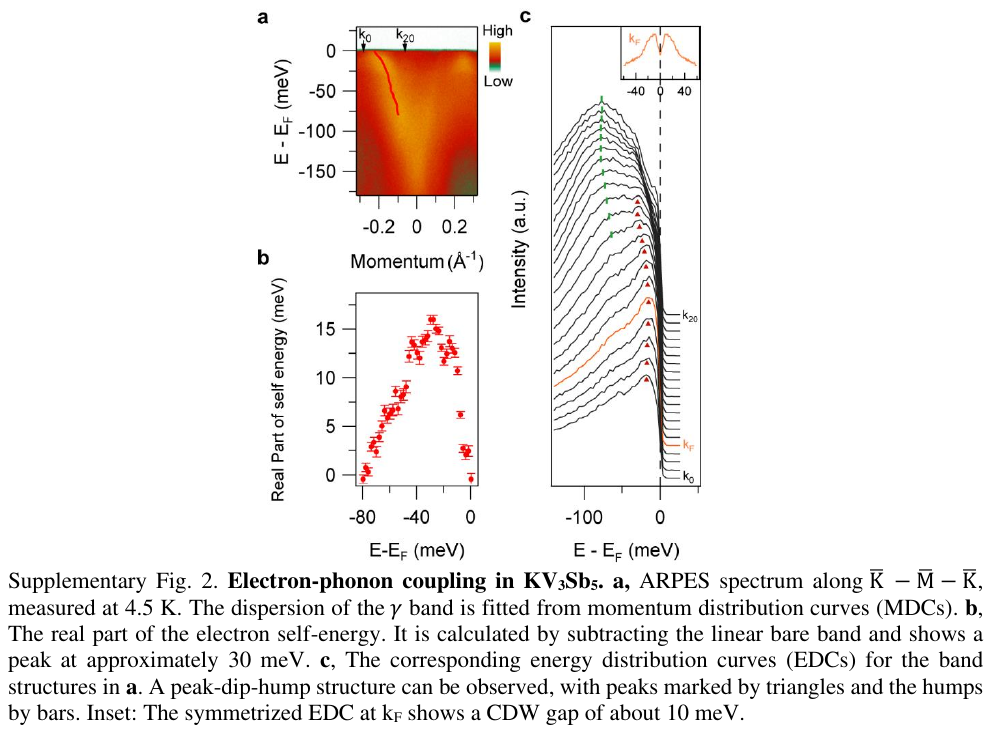}
\end{figure*}
\begin{figure*}[htbp]
\centering\includegraphics[width = 1\columnwidth,page=3] {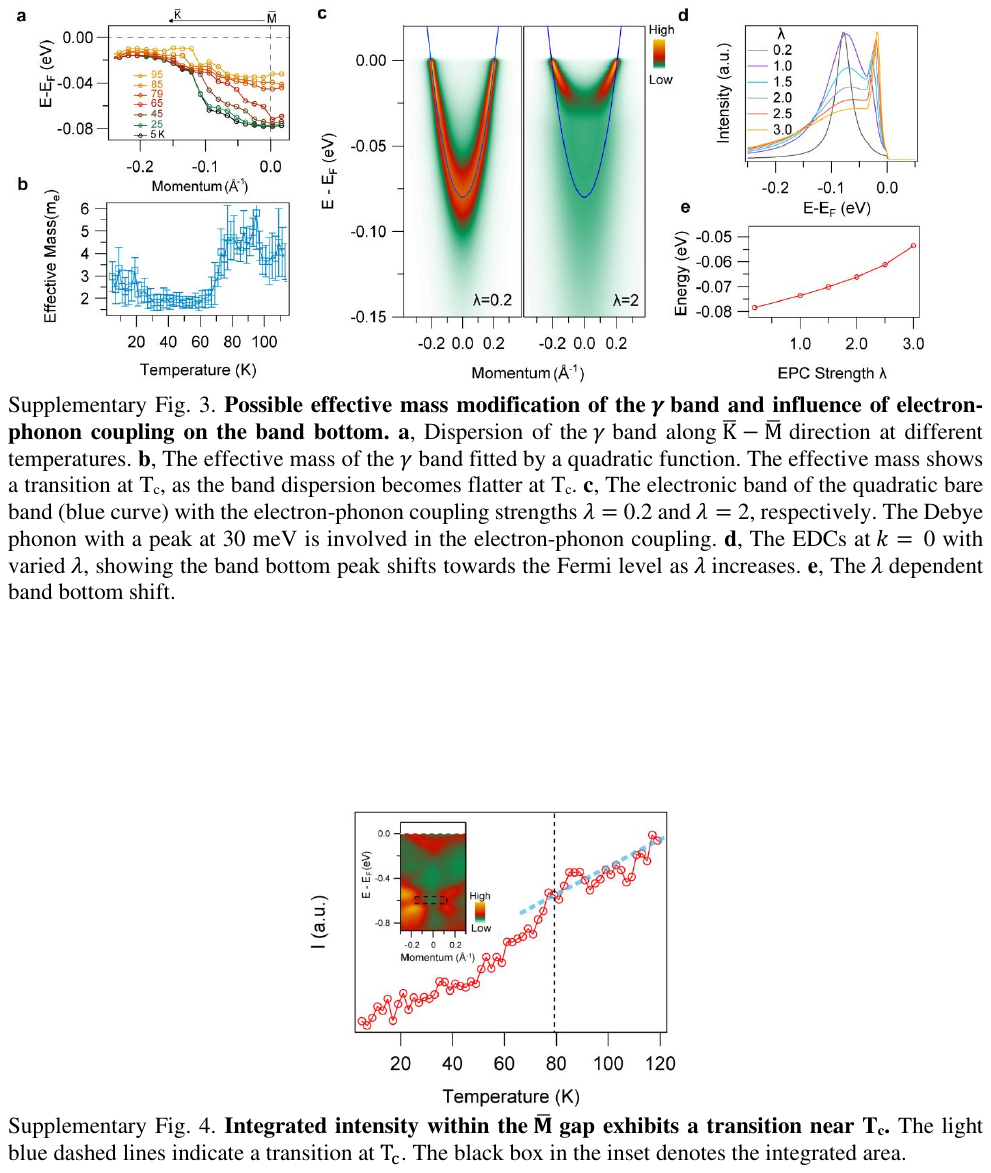}
\end{figure*}
\begin{figure*}[htbp]
\centering\includegraphics[width = 1\columnwidth,page=4] {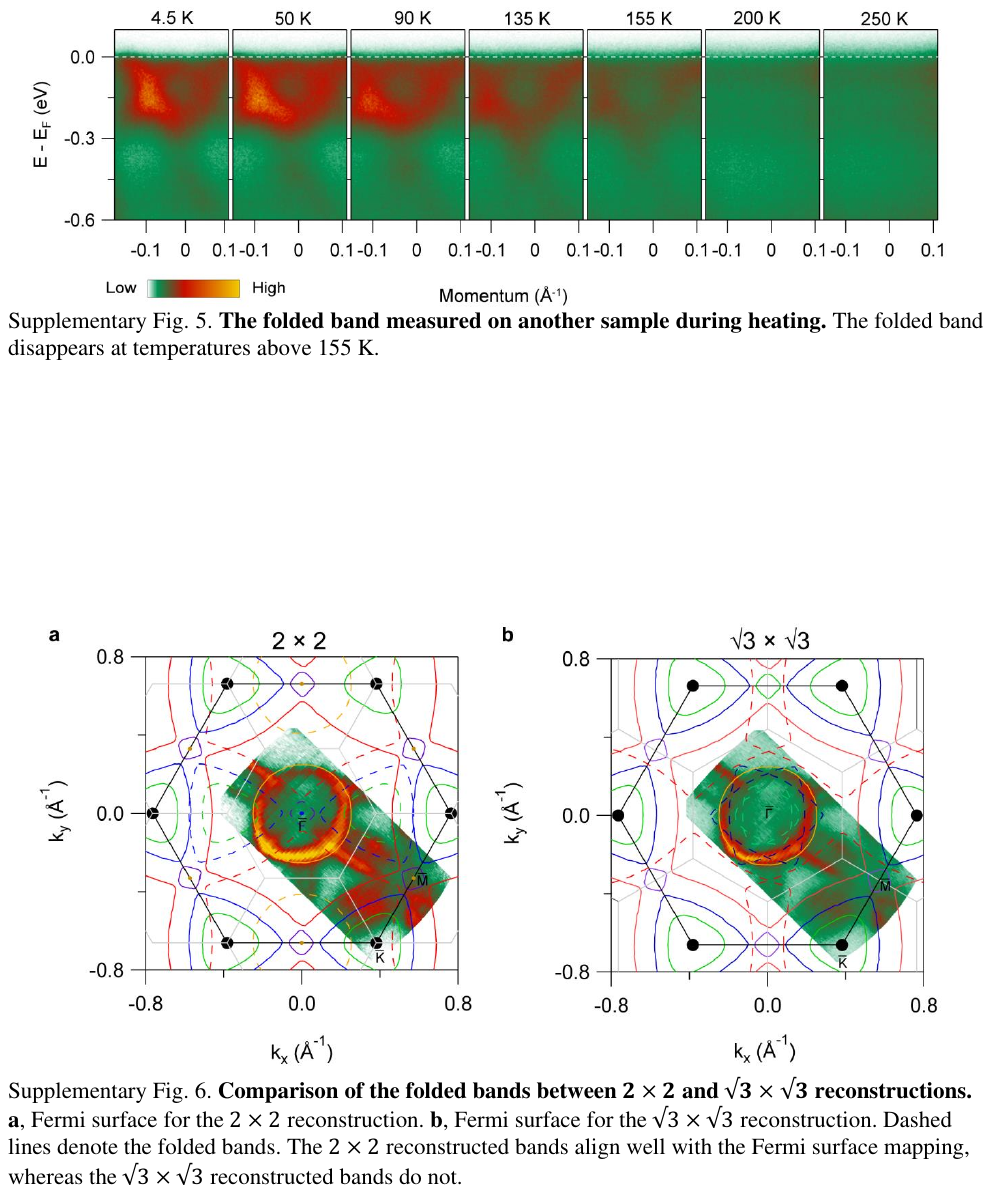}
\end{figure*}
\begin{figure*}[htbp]
\centering\includegraphics[width = 1\columnwidth,page=5] {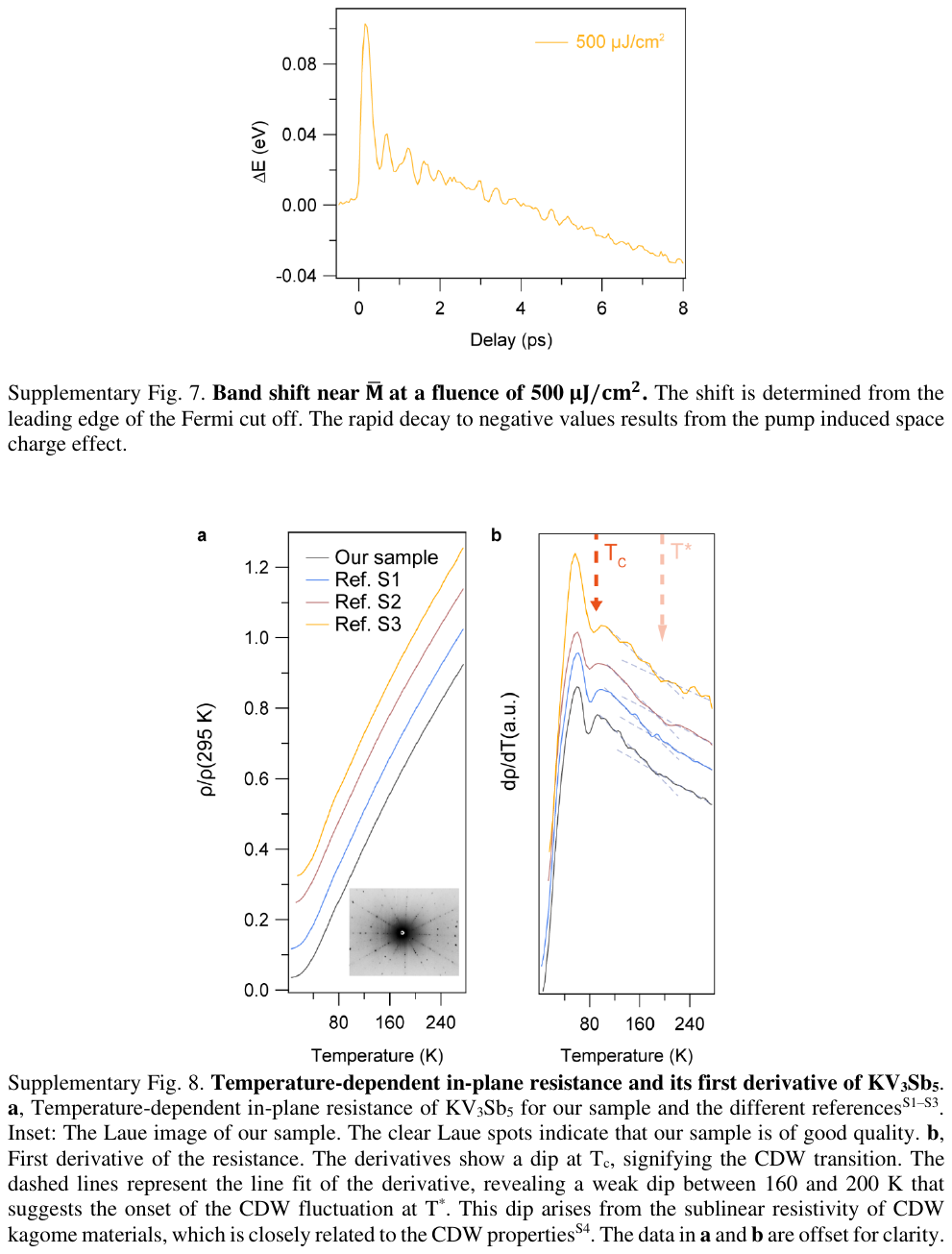}
\end{figure*}
\begin{figure*}[htbp]
\centering\includegraphics[width = 1\columnwidth,page=6] {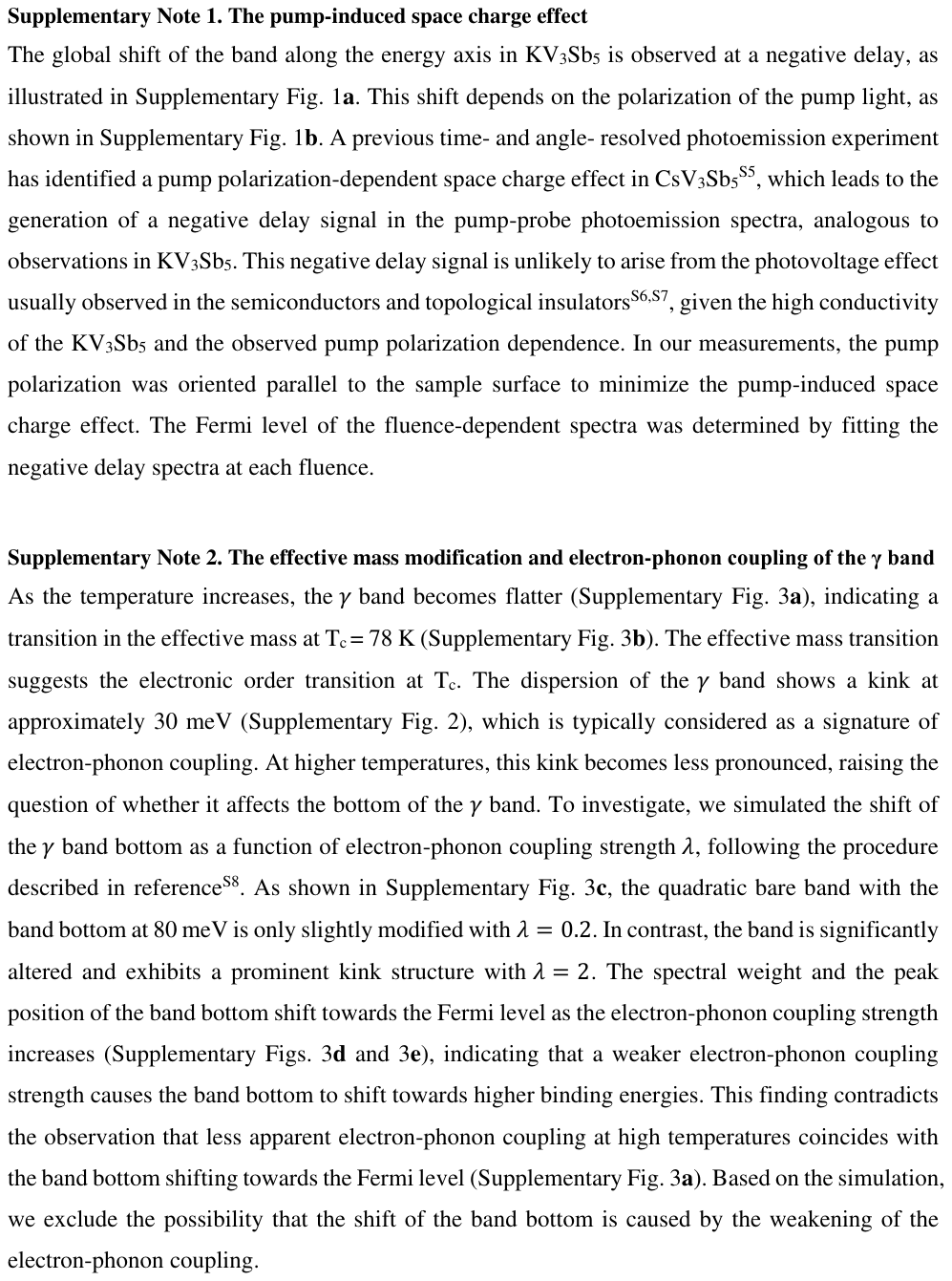}
\end{figure*}
\begin{figure*}[htbp]
\centering\includegraphics[width = 1\columnwidth,page=7] {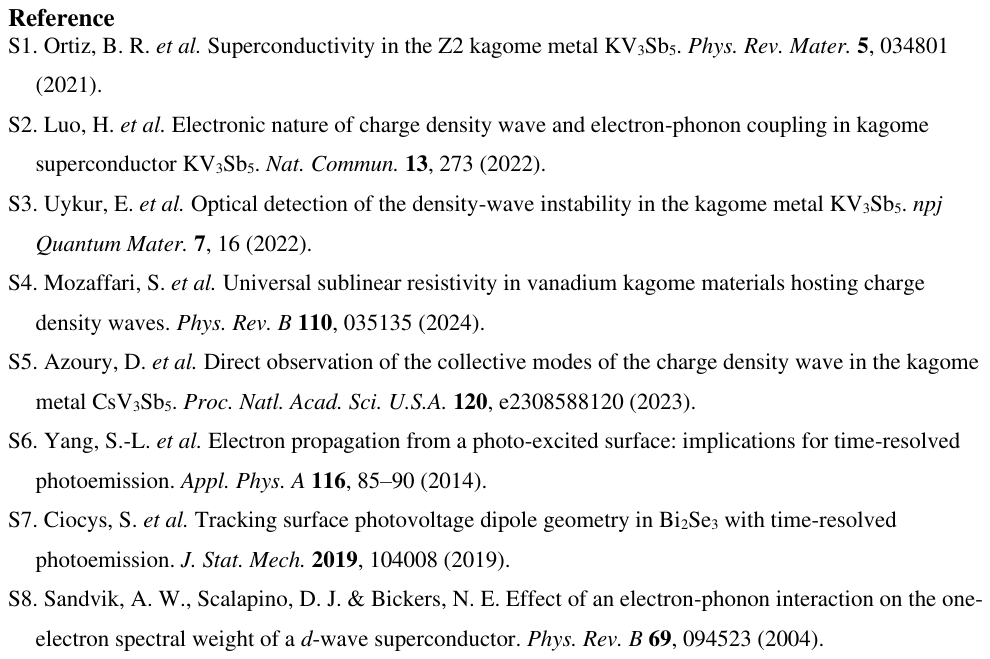}
\end{figure*}

\end{document}